\documentclass[twocolumn,twoside,11pt]{article}

\usepackage[left=0.58in,top=0.72in,bottom=0.7in,right=0.58in,headsep=0.28in,footskip=0.3in,footnotesep=4pt]{geometry}

\usepackage[para]{footmisc}
\usepackage[misc]{ifsym}

\usepackage{graphicx}
\graphicspath{{Figures/}}

\usepackage[colorlinks=true,citecolor=blue,linkcolor=blue,urlcolor=blue]{hyperref}

\usepackage{dcolumn}
\usepackage{xcolor,balance}
\usepackage{fancyhdr}
\usepackage{lipsum}
\usepackage{soul}
\usepackage{braket}

\usepackage{kbordermatrix}
\usepackage{amsmath,amssymb}
\usepackage{titlesec}
\usepackage{lmodern}
\usepackage{booktabs}
\usepackage{footnote}
\usepackage{lettrine}
\usepackage{cite}

\makeatletter 
\renewcommand{\fnum@figure}{\textbf{FIG.~\thefigure}}
\makeatother

\makeatletter
\def\bbordermatrix#1{\begingroup \m@th
  \@tempdima 4.75\p@
  \setbox\z@\vbox{%
    \def\cr{\crcr\noalign{\kern2\p@\global\let\cr\endline}}%
    \ialign{$##$\hfil\kern2\p@\kern\@tempdima&\thinspace\hfil$##$\hfil
      &&\quad\hfil$##$\hfil\crcr
      \omit\strut\hfil\crcr\noalign{\kern-\baselineskip}%
      #1\crcr\omit\strut\cr}}%
  \setbox\tw@\vbox{\unvcopy\z@\global\setbox\@ne\lastbox}%
  \setbox\tw@\hbox{\unhbox\@ne\unskip\global\setbox\@ne\lastbox}%
  \setbox\tw@\hbox{$\kern\wd\@ne\kern-\@tempdima\left[\kern-\wd\@ne
    \global\setbox\@ne\vbox{\box\@ne\kern2\p@}%
    \vcenter{\kern-\ht\@ne\unvbox\z@\kern-\baselineskip}\,\right]$}%
  \null\;\vbox{\kern\ht\@ne\box\tw@}\endgroup}
\makeatother

\usepackage{array}
\newcolumntype{L}{>{\centering\arraybackslash}m{2 cm}}
\newcolumntype{C}{>{\centering\arraybackslash}m{3.2 cm}}
\usepackage{multirow}
\usepackage{orcidlink}
\usepackage{changepage}

\newcommand{\periodafter}[1]{#1.}

\definecolor{customgreen}{rgb}{0.306,0.463,0.51}
\titleformat{\section}[display]
  {\sffamily\large\bfseries}{}{0pt}{\large}
\titlespacing*{\section}{0pt}{0pt}{2pt}
\titleformat{\subsection}[runin]
  {\normalfont\bfseries}{}{0pt}{\periodafter}

\makeatletter
\def\@fnsymbol#1{\ensuremath{\ifcase#1\or *\or \dagger\or dagger\or \mathsection\or \mathparagraph\or \|\or **\or \dagger\dagger \or \text{\Letter} \else\@ctrerr\fi}}
\makeatother

\newcommand{\corresauth}[2][9]{\renewcommand{\thefootnote}{\fnsymbol{footnote}}\footnote[#1]{#2}\renewcommand{\thefootnote}{\arabic{footnote}}}

\begin{document}
\onecolumn
%
%\begin{center}
%{\color{red}\textbf{CONFIDENTIAL DRAFT: PLEASE DO NOT DISTRIBUTE}}
%\end{center}

\renewenvironment{abstract}
{\begin{adjustwidth}{0pt}{133.7pt} \sffamily}
{\end{adjustwidth}}

\begin{titlepage}
\pagestyle{fancy}
\null\par
\vskip 9em
\setlength{\parindent}{0pt}
{\sffamily\Huge Emulating Quantum Interference with Generalized Ising Machines \par}
\vspace{2em}
{\sffamily Shuvro Chowdhury{\vspace{2em}} \orcidlink{https://orcid.org/0000-0002-6325-0790}\footnote[1]{\sffamily Department of Electrical and Computer Engineering, University of California, Santa Barbara, Santa Barbara, CA 93106, USA,}\textsuperscript{,}\corresauth[9]{\sffamily email:  schowdhury@ucsb.edu}, Kerem Y. Camsari \orcidlink{https://orcid.org/0000-0002-6876-8812}\footnotemark[1] and Supriyo Datta \orcidlink{https://orcid.org/0000-0001-8577-984X}\footnote[2]{Elmore Family School of Electrical and Computer Engineering, Purdue University, IN 47907, USA}}

\vspace{0.25in}

\begin{abstract}
The primary objective of this paper is to present an exact and general procedure for mapping any sequence of quantum gates onto a network of probabilistic \emph{p-bits} which can take on one of two values $0$ and $1$. The first $n$ p-bits represent the input qubits, while the other p-bits represent the qubits after the application of successive gating operations. We can view this structure as a Boltzmann machine whose states each represent a \emph{Feynman path} leading from an initial configuration of qubits to a final configuration. Each such path has a complex amplitude $\psi$ which can be associated with a complex energy. The real part of this energy can be used to generate samples of Feynman paths in the usual way, while the imaginary part is accounted for by treating the samples as complex entities, unlike ordinary Boltzmann machines where samples are positive. Quantum gates often have purely imaginary energy functions for which all configurations have the same probability and one cannot take advantage of sampling techniques. Typically this would require us to collect $2^{nd}$ samples which would severely limit its utility. However, if we can use suitable transformations to introduce a real part in the energy function then powerful sampling algorithms like Gibbs sampling can be harnessed to get acceptable results with far fewer samples and perhaps even escape the exponential scaling with $nd$. This algorithmic acceleration can then be supplemented with special-purpose hardware accelerators like Ising Machines which can obtain a very large number of samples per second through a combination of massive parallelism, pipelining, and clockless mixed-signal operation made possible by codesigning circuits and architectures to match the algorithm. Our results for mapping an arbitrary quantum circuit to a Boltzmann machine with a complex energy function should help push the boundaries of the simulability of quantum circuits with probabilistic resources and compare them with NISQ-era quantum computers.
\end{abstract}

\vfill

{\color{customgreen}\rule{\textwidth}{.3ex}}
\end{titlepage}

\date{\today}

%\pacs{}
%\maketitle
\twocolumn
\section{Introduction}
\label{sec:Intro}

\begin{figure}[!ht]
	\centering
	\vspace{0pt}
	\includegraphics[width=0.95\linewidth,keepaspectratio]{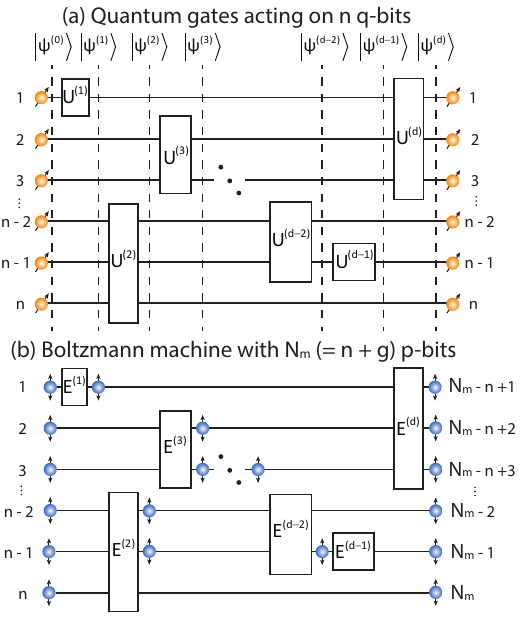}
	\caption{\textbf{Mapping between quantum circuits and Boltzmann machines:} (a) In quantum computing, a sequence of unitary gates ${U}^{(1)}$,${U}^{(2)}$,$\cdots$,${U}^{(d)}$ is applied to an initial wavefunction $\vert\psi^{(0)}\rangle$ and a final wavefunction $\vert\psi^{(d)}\rangle$ is obtained. A classical bitstream of 0s and 1s is then obtained from this wavefunction by ``measuring" the qubits (not shown) which leads to the collapse of the wavefunction and it is a non-reversible operation. (b) The quantum circuit in (a) is mapped into a Boltzmann machine with p-bits. Note that although the quantum gates in Fig.~\ref{fig:map}a are time-ordered, the p-bit network in Fig.~\ref{fig:map}b is a reciprocal BM described with an energy function which is obtained by modeling each $k$-th quantum gate via a corresponding complex energy function, $E^{(k)}$. Nevertheless, the time sequence of gates is reflected in the BM's energy function. While in the quantum case, the number of qubits remains the same, in the Boltzmann analog, the number of p-bit required increases with the number of applied gates $d$ as $n+\text{g}(d)$. See Section~\ref{sec:E} for more details on $\text{g}(d)$).}
	\label{fig:map}
\end{figure}

\lettrine[lines=3,findent=1pt,nindent=0em,lraise=0.05,loversize=0]{{\sffamily \textcolor{customgreen}{{Q}}}}{}uantum computing is based on the use of quantum gates to perform $d$ successive unitary transformations (gates) ${U}^{(1)}, {U}^{(2)}, \cdots, {U}^{(d-1)}, {U}^{(d)}$ on a set of $n$ qubits so that their wavefunctions evolve from an initial $\vert\psi^{(0)}\rangle$ to a final $\vert\psi^{(d)}\rangle$ (Fig.~\ref{fig:map}a) \cite{Nielsen2011}. Each of these wavefunctions $\vert\psi^{(d)}\rangle$ has $2^n$ components, coming from a tensor product of $n$ single qubit wavefunctions with two complex components each. In classical computing, a direct deterministic calculation requires us to multiply $2^n \times 2^n$ transformation matrices with exponentially large memory requirements as $n$ increases. By contrast, a quantum computer requires only $n$ qubits which naturally live in $2^n\times2^n$ Fock space.

Interestingly, a classical probabilistic computer too lives in this $2^n\times2^n$ Fock space, but is described by a probability distribution function with positive elements, unlike the complex wavefunction that describes a quantum computer. Indeed in the seminal paper that inspired the field of quantum computing, Feynman remarked \cite{Feynman1982} \emph{\ldots The only difference between a probabilistic classical world and the equations of the quantum world is that somehow or other it appears as if the probabilities would have to go negative \ldots} However, this fundamental difference leads to interesting properties of quantum circuits such as entanglement and interference \cite{Nielsen2011} that give quantum computing its theoretical power. %\cite{Horodecki2009review}, \cite{mermin2007quantum},

In the field of adiabatic quantum computing (AQC), it is well-known that a significant subset of Hamiltonians like the transverse field Ising model (TFIM) are \emph{stoquastic}, which means that the elements of the matrix $\exp{(-H)}$ are positive and can be evaluated using Monte Carlo techniques \cite{Troyer2005,Camsari2019} similar to those widely used to evaluate classical probabilities. It is well-known that quantum Monte Carlo (QMC) techniques can be used for \emph{non-stoquastic} matrices as well though their accuracy is limited by the \emph{sign problem}. This is important since gate-based quantum computing (GQC) is based on unitary matrices of the form $\exp{(-iH)}$ which are nearly always non-stoquastic.

The primary contribution of this paper is to present an exact and general procedure for mapping any sequence of quantum gates onto a probabilistic computer. This mapping should be particularly useful in view of the advent of special-purpose hardware accelerators known as ``Ising Machines'' and ``digital annealers'' \cite{yamaoka201520k,mcmahon2016fully,inagaki2016coherent,wang2017oscillator,chou2019analog,borders2019integer,dutta2019experimental,aramon_fujitsu} which are being used to simulate the statistical mechanics of Ising models, onto which many known combinatorial optimization problems have been mapped \cite{lucas2014ising}. These special purpose machines make use of random number generators (RNGs) to obtain a very large number of samples per second \cite{sutton2019autonomous} through massively parallel operation and can be used to speed up the emulation of quantum circuits, once they have been mapped onto a probabilistic framework.

Note that we expect the probabilistic computer to improve the time to solution through the prefactor, but we do not expect it to change the asymptotic scaling behavior. For example, we present results showing the implementation of Shor's algorithm in less than 2 hours on an ordinary laptop for a 27-bit number ($n=27$), which is much larger than previously reported. But the time scales exponentially $O(2^n)$ unlike a true noiseless quantum computer which is expected to show linear scaling $O(n)$. However, quantum computers require stringent control of phase which is difficult even at cryogenic temperatures \cite{Noh2020efficientclassical,King2021} and current research activities in quantum computing have moved to noisy intermediate scale quantum (NISQ) computing for which there is no established asymptotic scale-up. As such there is strong interest in pushing the boundaries of classical computing \cite{gottesman1998,Aaronson2004,Smith2019,zhou2020limits,carrasquilla2019probabilistic,napp2019efficient,noh2020efficient}. Probabilistic computers can be built to operate at room temperature using existing technology, and energy-efficient compact realizations may be possible using stochastic nanomagnets \cite{borders2019integer}.
%\cite{arute2019quantum,Zhong2020quantum,kim2023evidence,morvan2023phase} % ,pan2022solving,begusic2023fast

The basic procedure for translating a q-circuit into a p-circuit is shown in Fig.~\ref{fig:map}. The first $n$ p-bits represent the input qubits, while the other p-bits represent the qubits after the application of successive gating operations (Fig.~\ref{fig:map}b). In the spirit of earlier works \cite{VandenNest2009,Geraci_2010, De_las_Cuevas_2011,Bremner2016,Boixo2018,Fujii_2017,jnsson2018,pehle2020neuromorphic}, we can view this structure as a Boltzmann machine whose states each represent a \emph{Feynman path} leading from an initial configuration of qubits to a final configuration, via specific intermediate configurations. Each such path has a complex amplitude $\psi$ which can be associated with a complex energy. The real part of this energy can be used to generate samples of Feynman paths in the usual way, while the imaginary part is accounted for by treating the samples as complex entities, unlike ordinary Boltzmann machines where samples are positive.

Recently, Boltzmann machines with complex weights have also been employed with machine learning techniques for various classes of quantum Hamiltonians. It has been shown that certain representations of Boltzmann machines can be trained to obtain the ground state of a Hamiltonian, and can even be used to represent quantum states exactly \cite{Carleo602,Carleo2018,Gao2017,Carrasquilla2019}. Our approach here is fundamentally different in that no training of weights is involved. Our weights are obtained analytically as described in Section~\ref{sec:E} and might be called ``one-shot'' learning.

\subsection{Organization of the paper} 
In Section~\ref{sec:E} we describe how we obtain our rules for translating one qubit and two-qubit gates into a complex energy function $E$ governing the corresponding complex Boltzmann machine, using the Feynman path approach \cite{Boixo2018}. In Section~\ref{sec:MCSampling}, we discuss probabilistic sampling from the Feynman paths and show, how the existing Ising machines with some modification should be able to perform such sampling. In Section~\ref{sec:IllustrateHadamard}, we present an illustrative example for a quantum circuit with large depth but only a single qubit while in Section~\ref{sec:Shor}, we discuss a shallow circuit with many qubits. Finally in Section~\ref{sec:Hardware}, we briefly discuss the possibility of orders of magnitude hardware acceleration by mapping our algorithm onto special-purpose classical circuits consisting of interconnected p-bits analogous to the interconnected qubits that comprise a quantum processor.

\section{Complex Energy Functions for Quantum Gates}
\label{sec:E}

The basic rule for writing a complex energy function for a sequence of gate operations can be obtained as follows. First, we note that the elements of the overall transformation matrix are given by a matrix product of the individual matrices
\begin{equation}
	U_{\alpha,\beta} = \sum_{p_1, \cdots , p_{d-1}} U^{(d)}_{\alpha,p_{d-1}} U^{(d-1)}_{p_{d-1},p_{d-2}} \cdots  U^{(2)}_{p_{2},p_{1}} U^{(1)}_{p_{1},\beta}
	\label{eq:U}
\end{equation}
where $\vert\beta\rangle$ and $\vert\alpha\rangle$ correspond to the initial and the final state of the qubits respectively. We express each $k$\textsuperscript{th} element of the transformation matrices in terms of a corresponding energy function, which in general can be complex:
\begin{eqnarray}
	U^{(k)}_{p,q} &=& e^{-E^{(k)}_{p,q}} \hspace{0.25in} \Rightarrow E^{(k)}_{p,q} = -\ln{\left(U^{(k)}_{p,q}\right)}.
	\label{eq:UE}
\end{eqnarray}
Using (\ref{eq:UE}) in (\ref{eq:U}), we can write
\begin{eqnarray}
	U_{\alpha,\beta} &=& \sum_{p_1, \cdots , p_{d-1}} e^{ -\big(E^{(d)}_{\alpha,p_{d-1}} + E^{(d-1)}_{p_{d-1},p_{d-2}} + \cdots + E^{(2)}_{p_{2}, p_{1} } + E^{(1)}_{p_{1},\beta } \big) }\nonumber \\
	&=&  \sum_{p_1, \cdots , p_{d-1}} e^{-\Re(E) } \ e^{ -i \Im(E) }
	\label{eq:path}
\end{eqnarray}
where ${\Re(E) }$ and ${\Im(E) }$ represent the real and imaginary parts of the total energy function $E$ obtained by summing the individual ones
\begin{equation}
	E (p_1, \cdots , p_{d-1})= E^{(d)}_{\alpha,p_{d-1}} + E^{(d-1)}_{p_{d-1},p_{d-2}} + \cdots + E^{(1)}_{p_{1},\beta}.
\end{equation}

\noindent Equation~(\ref{eq:path}) is an exact result, but it represents a sum over a large number of paths (also referred to as Feynman paths in this context)
\[ \beta\rightarrow p_1\rightarrow p_2\rightarrow \cdots \rightarrow p_{d-1}\rightarrow \alpha\]
(from an initial state $\beta$ to a final state $\alpha$) which grows exponentially in $\textit{n} $.

Usually, the energies are real so that $e^{-E}$ can be interpreted as a probability, and probabilistic approaches allow us to sample the most important paths based on powerful algorithms like Metropolis or Gibbs sampling. For complex energies, we can still sample the paths based on the real part of $E$ while the imaginary part can be interpreted as the complex contribution of unit magnitude contributed by a particular path. Next, we will show how to obtain energy functions for one and two-qubit gates and the sampling procedure will be discussed in the next section.

\subsection{One-qubit gates} 
Any one-qubit gate is in general described by a transformation matrix of the form
\begin{equation}
	{U}^{(d)}=\kbordermatrix{
		&|0^{(d-1)}\rangle & |1^{(d-1)}\rangle\\
		\langle 0^{(d)}|&a_1& b_1 \\
		\langle 1^{(d)}|&c_1 & A_1}.
\end{equation}

Using (\ref{eq:UE}) we can immediately write the energy function in tabular form
\begin{eqnarray}
	E^{(d)} =\quad &&\kbordermatrix{
		&s^{(d-1)}=0 & s^{(d-1)}=1\\
		s^{(d)}=0&-\ln{(a_1)} & -\ln{(b_1)}\\
		s^{(d)}=1&-\ln{(c_1)} & -\ln{(A_1)}}. \nonumber
\end{eqnarray}

It is straightforward to convert this tabular result into a Boolean sum of products expression  \cite{wakerly2016digital}
\begin{equation}
	\begin{split}
		E^{(d)} =  & - (1-s^{(d)} ) (1-s^{(d-1)} ) \ \ln{(a_1)}  \\
		&- \ (1-s^{(d)} ) \ s^{(d-1)} \ \ln{(b_1)} \\
		&- \ s^{(d)} \ (1-s^{(d-1)} ) \ \ln{(c_1)} \\
		&- \ s^{(d-1)} \ s^{(d)} \ \ln{(A_1)}
	\end{split}
	\label{eq:energyBoolean}
\end{equation}

\noindent which simplifies to
\begin{eqnarray}
	E^{(d)} = &-& \ln{(a_1)}+   s^{(d)} \ \ln{(a_1/c_1)}  + s^{(d-1)} \ \ln{(a_1/b_1)} \  \  \  \  \nonumber\\
	& +& s^{(d-1)} s^{(d)} \  \ln{(b_1c_1/a_1 A_1)}. 
	\label{eq:EB}
\end{eqnarray}
Note that this energy function has linear and quadratic terms corresponding to one-body and two-body interactions in an Ising model, which requires a Boltzmann machine with a linear synaptic function derived from the gradient of the energy function. 

\subsection{Two qubit gates} 
Any two-qubit gate is in general described by a transformation matrix of the form
\begin{equation}
	\footnotesize
	{U}^{(d)}=\kbordermatrix{
		&|00^{(d-1)}\rangle & |10^{(d-1)}\rangle & |01^{(d-1)}\rangle & |11^{(d-1)}\rangle\\
		\langle 00^{(d)}| &a_1&b_1 &a_2 &b_2\\
		\langle 10^{(d)}| &c_1 &A_1 &c_2 &A_2 \\
		\langle 01^{(d)}| &a_3 &b_3 &a_4 &b_4 \\
		\langle 11^{(d)}| &c_3 &A_3 &c_4 &A_4}.
\end{equation}
\noindent Using (\ref{eq:UE}) we can immediately write the energy function in tabular form
\begin{equation}
	E^{(d)}=\footnotesize\begin{array}{cc}
		& s_1^{(d-1)}s_2^{(d-1)}\\
		\rotatebox{90}{\hspace{-0.4cm}$s_1^{(d)}s_2^{(d)}$} &
		\kbordermatrix{
			&00 & 10 & 01 & 11\\
			00 &-\ln{(a_1)}& -\ln{(b_1)} &-\ln{ (a_2)} & -\ln{(b_2)}\\
			10 & -\ln{(c_1)} & -\ln{(A_1)} & -\ln{(c_2)} & -\ln{(A_2)} \\
			01 & -\ln{(a_3)} & -\ln{(b_3)} & -\ln{(a_4)} & -\ln{(b_4)} \\
			11 & -\ln{(c_3)} & -\ln{(A_3)} & -\ln{(c_4)} & -\ln{(A_4)} }
	\end{array}
	\label{eq:ET2}
\end{equation}
\noindent \normalsize Once again this tabular result can be translated into a Boolean function like (\ref{eq:EB}), but the energy function will have three-body and four-body interactions whose gradient leads to non-linear synaptic terms. Equation (\ref{eq:ET2}) with such three-body and four-body terms represents a higher-order Ising model that has been discussed in the learning context by  Ref.~\cite{sejnowski1986}. Such ``generalized'' Ising models can be solved much like ordinary Ising models with two-body interactions, provided that the synaptic feedback can be computed, for example by an FPGA \cite{mcmahon2016fully}. Naturally, the resistive crossbar arrays that accelerate \emph{linear} synaptic operations \cite{hu2016dot} would not be suitable for this purpose. 

It is possible to eliminate the three-body and four-body terms at the expense of additional p-bits by decomposing the gates in terms of a sequence of rotation operations \cite{Kim2000,VandenNest2009}, making the synaptic function linear. Alternatively, the three and four-body interactions can be reduced to standard 2-body interactions using the methods described in \cite{jiang2018quantum,biamonte2008nonperturbative, tanburn2015reducing,dattani20194variable}. 

In general, in this scheme, it would require two p-bits (one input p-bit and one output p-bit) to implement each single qubit gate. If the quantum circuit consists of $d$ one qubit gates in series then in Fig.~\ref{fig:map}, $\text{g}(d)$ would be $d$. But if the one qubit gate is diagonal (like the phase gates) then one can implement that gate using just a single p-bit (same p-bit will represent both input and output) (see the appendix) and $\text{g}(d)$ would be 0 for a quantum circuit with `d' such gates in series. Similarly, for two-qubit gates one would require 4 p-bits and $\text{g}(d)$ here would be $2d$ for a circuit consisting of only two-qubit gates in series whenever it is possible to implement 4-body interactions without inserting any intermediate p-bits. For diagonal gates (like controlled phase gates), once again $\text{g}(d)$ would be 0 since those gates can be implemented using just two p-bits. For a general quantum circuit with $n$ qubits, $\text{g}(d)$ would be approximately a linear function of $n$.

It is an important result in quantum computing that single qubit gates along with CNOT gates are universal for quantum computation \cite{Nielsen2011}. Therefore, the methodology presented here for translating one and two qubits gates can in principle be used to translate any quantum circuit into a p-bit network although it is possible to extend the methodology for gates operating on more than two qubits directly (for example,  Toffoli or Fredkin gates which operate on three qubits).

Finally, we also note that it is not necessary to turn the tabular results in (\ref{eq:EB}) and (\ref{eq:ET2}) into a Boolean function, one can store these tabular values in a lookup table and look for them up from that table as necessary when summing to get the total energy of a path and then use a Metropolis algorithm based sampling approach with the help of a ``dedicated kernel'' as explained in the next section.

\subsection{Adiabatic Quantum Computing}

Before moving on, we want to note that in this paper we focus on GQC which is based on unitary transformations ${U}\propto \exp{(-i{H}})$. In this case, the energy functions are rarely real since that requires the elements of the ${U}$ matrix to be all real and positive, which is seldom the case. However, our approach is also applicable to adiabatic quantum computing (AQC) based on $\exp{(-\beta {H})}$ which can often have purely real and positive elements leading to purely real energy functions. Such Hamiltonians are classified as stoquastic (see for example, \cite{bravyi2006complexity}) and can be emulated with standard BM's. Non-stoquastic Hamiltonians requiring complex BM's form a special subset of all problems of interest in AQC, while in GQC many problems belong to this category. 

Interestingly, in one respect GQC is simpler than AQC because the GQC evolution operators are naturally built out of the product of few (typically 1 and 2) qubit operations.
\begin{equation}
	{U} =   \exp{\left(-i {H_1}\right)} \exp{\left(-i {H_2}\right)} \cdots \nonumber
\end{equation}
By contrast, it is not straightforward to do the reverse, namely to break up the evolution operator $\exp{\left(-\beta {H}\right)}$ for AQC into a product of separate terms corresponding to the components ${H}={H_1} + {H_2} + \cdots $, since $ \exp{\left(-\beta {H}\right)} \neq   \exp{\left(-\beta {H_1}\right)} \exp{\left(-\beta {H_2}\right)} \cdots \nonumber$ \noindent unless the components ${H_1},{H_2}, \cdots$ commute. The standard approach is the Suzuki-Trotter transformation \cite{Suzuki1976} which breaks up ${H}$ into $r$ replicas by writing
\begin{eqnarray}
	&& \exp{\left(-\beta {H}\right)} =   \left[\exp{\left(-\beta {H} / r\right)}\right]^{r}   \nonumber \\
	&& \approx \big( \exp{\left(-\beta {H_1}/r\right)}  \exp{\left(-\beta {H_2}/r\right)} \cdots \big)^{r} \nonumber 
\end{eqnarray}
assuming that $r$ is large enough to make the commutators of ${H_1}/r$, ${H_2}/r, \cdots$ negligible. This paper focuses on GQC where replicas are not needed as explained above.

\section{Sampling from Feynman paths}
\label{sec:MCSampling}

Note that for a given input state $\beta$ and output state $\alpha$ our objective is to evaluate $U_{\alpha,\beta}$ from (\ref{eq:U}) representing the summation of a very large number of Feynman paths each of which can be visualized as a state of the Boltzmann machine. However, unlike ordinary Boltzmann machines, each term is complex, requiring a modification of the standard sampling techniques as summarized below.

Monte Carlo techniques start by rewriting (\ref{eq:U}) in the form
\begin{equation}
	U_{\alpha,\beta} = \sum_{\{C\}} {p_{C} W_{C}}
	\label{eq:U343}
\end{equation}
where we have defined
\begin{equation}
	W_C = \cfrac{U^{(d)}_{\alpha,p_{d-1}} U^{(d-1)}_{p_{d-1},p_{d-2}} \cdots U^{(2)}_{p_{2},p_{1}}U^{(1)}_{p_{1},\beta}}{p_C}.
	\label{eq:U344}
\end{equation}
$\{C\}$ is the set of all possible paths (or configurations) from $\beta\to\alpha$ 
\begin{equation}
	C:\beta\rightarrow p_1\rightarrow p_2\rightarrow \cdots \rightarrow p_{d-1}\rightarrow \alpha \nonumber
\end{equation}
and $p_{C}$ is a probability assigned to the path $C$ as described below. The idea is to approximate the exact sum in (\ref{eq:U343}) with a sum over $T$ sample configurations generated according to the probability distribution $\{p_{C}\}$.
\begin{equation}
	U_{\alpha,\beta} \approx \cfrac{1}{T}\sum_{k=1}^T  W_{C_k}
	\label{eq:U23}
\end{equation}
where $C_k$ is the configuration sampled in $k$\textsuperscript{th} sample. The average estimate from many sampled sums (\ref{eq:U23}) will equal the correct value from the exact sum in (\ref{eq:U}) or (\ref{eq:U343}).

Following the central limit theorem, the standard deviation $\text{std}(T)$ of the Monte Carlo estimator with $T$ samples is given by
\begin{equation}
	\text{std}(T) = \cfrac{\text{std}(\{W_C\},\{p_C\})}{\sqrt{T}}
	\label{eq:U26}
\end{equation}
where
\begin{equation}
	\text{std}(\{W_C\},\{p_C\}) = \left(\sum_{\{C\}}{p_C(W_C-U_{\alpha,\beta})^2}\right)^{1/2}
	\label{eq:U30}
\end{equation}
Sampling techniques seek to minimize any $\text{std}(\{W_C\},\{p_C\})$  through a judicious choice of $\{p_C\}$.

\begin{figure}[!t]
	\centering
	\vspace{0pt}	\includegraphics[width=0.8\linewidth,keepaspectratio]{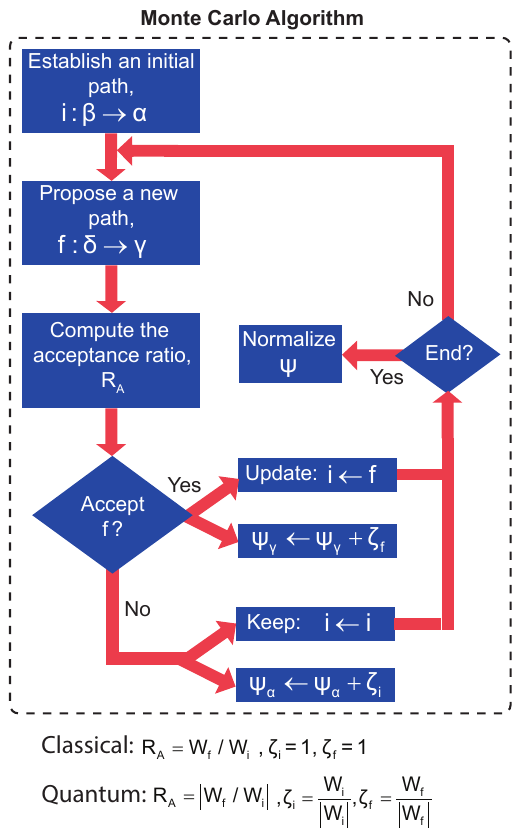}
	\caption{\textbf{Monte Carlo procedure for sampling from Feynman paths}: starts with an initial configuration or \emph{path} ($i$: a path from input state $\beta$ to output state $\alpha$). \textbf{Step 1:} Propose a new configuration ($f$: another path from input state $\delta$ to output state $\gamma$) based on some scheme like Gibbs sampling. \textbf{Step 2:} Accept or reject the proposal based on the ratio of $\exp{(-E)}$ for the proposed and initial configurations, using an algorithm like Metropolis-Hastings or Glauber. In classical Monte Carlo, E is real and $\exp{(-E)}$ is a positive number. In quantum problems, E can be complex, and we use the absolute value of $\exp{(-E)}$ as the acceptance ratio. \textbf{Step 3:} Add result to the output bin. In classical Monte Carlo, the output bin corresponding to the end state of the sampled path ($\psi_{\gamma}$ if the proposed path is accepted or $\psi_{\alpha}$ if the proposal is rejected) is just incremented by 1. In order to include the ``sign-correction" in quantum problems, the ``phase" of the path is added instead of $+1$. This loop continues until a sufficient number of samples are taken. \textbf{Step 4:} Finally the output wavefunction is normalized by dividing each bin with the square root of the sum of the squares of values of all bins (also known as $\text{L}_2$ normalization).}
	\label{fig:MCBlock}
\end{figure}

Classically it is common to choose 
\begin{equation}
	p_{C}\propto U^{(d)}_{\alpha,p_{d-1}} U^{(d-1)}_{p_{d-1},p_{d-2}} \cdots U^{(1)}_{p_{1},\beta}. \nonumber
\end{equation} A natural extension to quantum systems with complex path amplitudes is to choose  
\begin{equation}
	p_{C}\propto \big|U^{(d)}_{\alpha,p_{d-1}} U^{(d-1)}_{p_{d-1},p_{d-2}} \cdots U^{(1)}_{p_{1},\beta} \big|=\exp{(\Re{(E)})}
	\label{eq:weightedS}
\end{equation}
This choice leads to the flowchart shown in Fig.~\ref{fig:MCBlock}. The overall structure is very similar to the algorithms implemented in the modern digital annealers today \cite{aramon_fujitsu}. There are two minor but profound differences though: first, since the energies are complex, one should use the ratios of the absolute values of the weights (proportional to $\exp{(-\Re{(E)})}$) when computing the acceptance ratio. This is different from standard Boltzmann machines where each path carries a positive real weight (there the imaginary part of $E$ is always zero) which can be interpreted as a quantity directly proportional to the probability of that path. Second when counting the frequency of occurrence of a particular path (or configuration) one needs to add the phase of the weight of the path (which is equal to $\exp{(-i \Im{(E)})}$) instead of just adding +1. So one also needs to have the ability to add complex numbers which should be relatively straightforward and easy to incorporate. For positive and real weights as in standard Boltzmann machines, instead of adding the phase of the weight of the path, one always adds $+1$ which can also be viewed as each path having a zero phase associated with it, as shown in Fig.~\ref{fig:MCBlock}. 

\section{EXAMPLE 1: Single qubit, deep circuit}
\label{sec:IllustrateHadamard}

\begin{figure}[!t]
	\centering
	\vspace{0pt}
	\includegraphics[width=0.95\linewidth,keepaspectratio]{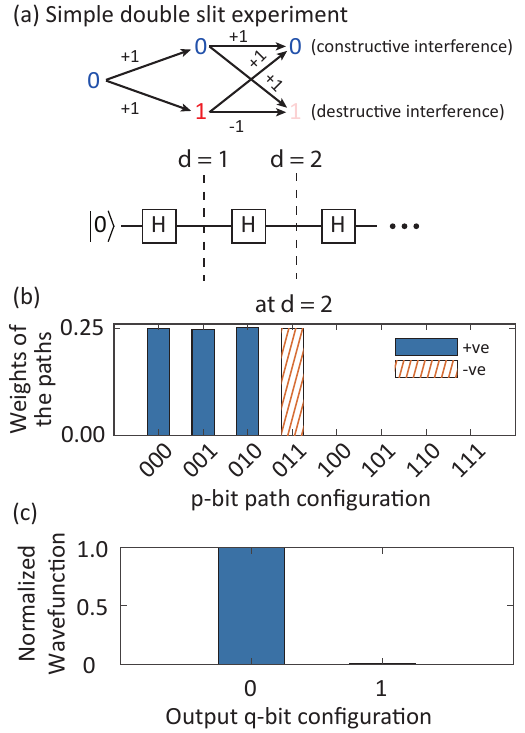}
	\caption{\textbf{Quantum interference with probabilistic sampling:} (a) One qubit with a string of Hadamard (H) gates applied in sequence. The input is clamped to $\vert 0 \rangle$ which can evolve into different output states $\vert 0 \rangle$ and $\vert 1 \rangle$ through different intermediate paths. (b) The weights of each of the four possible paths at $d=2$  are shown (the other four paths are not visited because those require the input to be set at $\vert 1\rangle$).  All paths contribute equally in terms of magnitude but the 011 path has a phase (hatch filled) which is opposite to that of the other three paths. (c) Normalized wavefunction plotted against different outputs which is what is expected from a direct matrix multiplication.}
	\label{fig:Had}
\end{figure}

It may seem surprising that complex probabilities can be included in a Monte Carlo simulation, but the basic idea can be appreciated with a one-qubit example. Consider one qubit driven by a sequence of Hadamard gates (Fig.~\ref{fig:Had}a), each represented by a transformation connecting qubits  in planes $d-1$ and $d$:
\begin{equation}
	{U}_{\text{Hadamard}}=\left(\frac{1}{\sqrt 2}\right) \kbordermatrix{
		&|0^{(d-1)}\rangle & |1^{(d-1)}\rangle \\
		\langle 0^{(d)}| &+1 &+1 \\
		\langle 1^{(d)}| &+1 &-1 }
	\label{eq:Had}
\end{equation}
Two applications of the gate result in an identity transformation:
\begin{equation}
	\big( {U}_{\text{Hadamard}} \big)^{2}=\kbordermatrix{
		&|0^{(d-1)}\rangle & |1^{(d-1)}\rangle \\
		\langle 0^{(d)}| &1 &0 \\
		\langle 1^{(d)}| &0 &1 }
\end{equation}
If the qubit is initialized to $\vert 0 \rangle$, then after one gate it has an equal probability of being either in $\vert 0 \rangle$ or $\vert 1 \rangle$, but after two gates the qubit is back to the $\vert 0 \rangle$ state with $100 \%$ probability.

This can be viewed as a very simple illustration of the classic double slit interference that Feynman used in his lectures to illustrate the difference between quantum and classical, or between electrons and bullets as he put it \cite{FeynmanBook}. Our Boltzmann machine emulates this quantum interference using a complex energy function given by
\begin{equation}
	E = i \pi (s_1 s_2 + s_2 s_3) + \mathrm{constant}
	\label{eq:eq2E}
\end{equation}
where $s_1, s_2, s_3$ represent the initial state,  the state after one gate, and the state after two gates respectively. It is straightforward to check that with $a_1 = b_1=c_1=-A_1=1$ in (\ref{eq:energyBoolean}), we obtain the result stated above in (\ref{eq:eq2E}). Starting from $\vert 0 \rangle$, the qubit will also end up in $\vert 0 \rangle$ after the two Hadamard gates. But the bullet starting from a ``0'' (corresponding to $s_1=0$) can end up in each of the final states ``0'' (corresponding to $s_3 = 0$) and ``1'' (corresponding to $s_3 = 1$) via two paths (each path is denoted by $s_1\rightarrow s_2\rightarrow s_3$) as follows

\begin{eqnarray}
	0&\rightarrow  0 \rightarrow 0   \qquad \text{(path 1)} \nonumber \\
	0&\rightarrow  1 \rightarrow 0   \qquad \text{(path 2)} \nonumber \\
	\text{or,}\qquad&\nonumber\\
	0&\rightarrow  0 \rightarrow 1   \qquad \text{(path 1)} \nonumber \\
	0&\rightarrow  1 \rightarrow 1   \qquad \text{(path 2)} \nonumber
\end{eqnarray}

All paths have the same $\Re(E)$ and appear with equal probability in the sampling process giving the same result for both outputs if the imaginary part is ignored as shown in Fig.~\ref{fig:Had}b. But the complex Boltzmann machine weights each path according to $\exp{(-i \Im{(E)})}$ so that the total contribution is
\begin{eqnarray}
	&& \text{Final  state 0: }  \rightarrow  e^{-i\pi(0 + 0)} +  e^{-i\pi(0 + 0)}  = 2 \nonumber \\
	&& \text{Final  state 1: } \rightarrow  e^{-i\pi(0 + 0)} +  e^{-i\pi(0 + 1)}  = 0. \nonumber
\end{eqnarray}

Our complex Boltzmann machine thus emulates quantum interference using classical bullets by associating complex energies with the paths taken by the bullet and hence in principle provides results that are exact when all possible paths taken by a bullet (between a given input state to a given output state) are considered into account (Fig.~\ref{fig:Had}c).

\begin{figure}[!t]
	\centering
	\vspace{0pt}
	\includegraphics[width=0.95\linewidth,keepaspectratio]{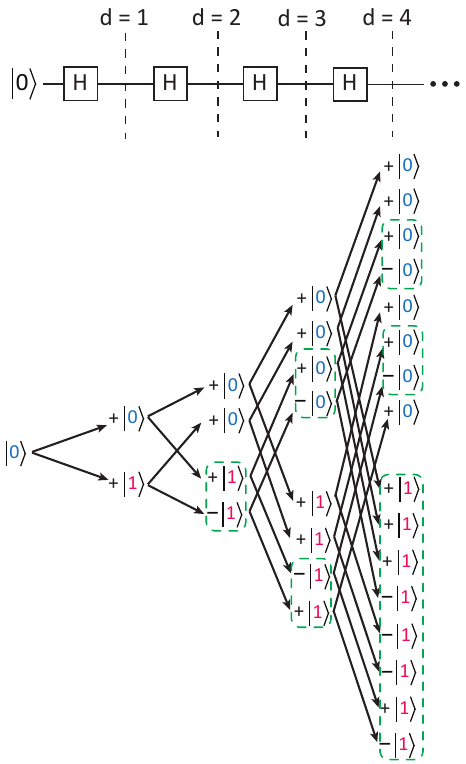}
	\caption{\textbf{Path cancellations in Hadamard chain:} Cancellation of paths are illustrated with a string of Hadamard (H) gates applied in sequence to an input clamped to $\vert 0 \rangle$. Half of the total paths cancel each other $\vert 1 \rangle$. Paths that cancel each other have been shown in the green dashed box. Even at $\vert 0 \rangle$, there are cancellations of paths, the amount of which also increases with the length of the chain. Only a small fraction of paths survive ($2^{d/2}$) out of total $2^{d}$ paths which after normalization produces a peak at $\vert 0 \rangle$.}
	\label{fig:HadChain}
\end{figure}

But p-bits with complex phases are still fundamentally different from qubits. We need four samples in order to get a net signal of 2 in state 0, and none in state 1. With noiseless qubits, two samples would have given us the same result, both would land in state 0. This difference becomes more pronounced as the depth of the circuit is increased (Fig.~\ref{fig:HadChain}), as we will now discuss.

\subsection{Sign problem in deep circuits} 

With two Hadamard gates the height of the correct peak is $1/2$; with $d$ ($d$ being even) gates it can be shown that the peak height is $1/2^{d/2}$  which becomes very small as d is increased. This means that many more samples will be needed to reduce the noise to an acceptable level.

We can estimate the noise using (\ref{eq:U26}) and (\ref{eq:U30}) as follows. With an even number of gates, $d$, there are $2^d$ paths, half of which reach the correct output and half reach the wrong output. Let us consider the wrong output with a net signal $U_{\alpha,\beta}$ = 0. Since the paths have weight $\pm1/2^{d/2}$, all with the same magnitude, it is best to use a uniform probability for all paths, with $p_{C}=2/2^{d}$, so that from (\ref{eq:U344}), $W_{C}=\pm2^{d/2}/2$, and from (\ref{eq:U26}) 
\begin{equation}
	\text{std(T)} =  \cfrac{2^{d/2}}{\sqrt{2T}}
	\label{eq:stdT}
\end{equation}
\noindent noting that only half the samples, $T/2$, reach the wrong output. To reduce this noise below the signal $1/2^{d/2}$, the number of samples T must exceed $2^d$.

\subsection{``Taming" the sign problem}

\begin{figure}[!ht]
	\centering
	\vspace{0pt}
	\includegraphics[width=0.95\linewidth,keepaspectratio]{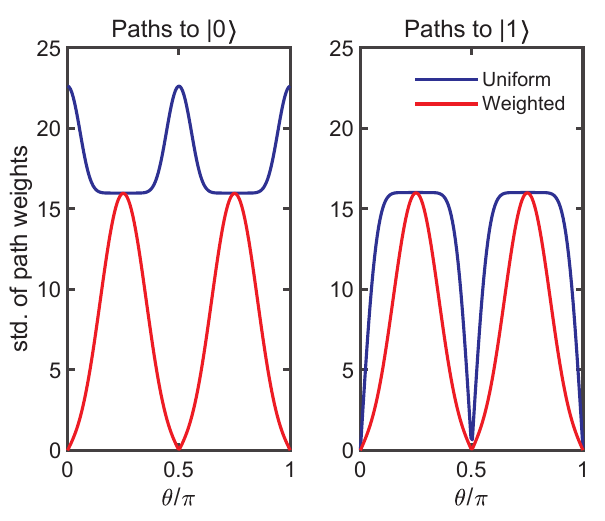}
	\caption{\textbf{Rotated Hadamard gate:} The variation of the prefactor of Eq.~(\ref{eq:U26}) vs. various rotational angles has been shown for $d=10$ rotated Hadamard gates in cascade. Weighted sampling yields lower prefactors and hence relatively lower number of samples will be required. The usual Hadamard gate corresponds to $\theta = \pi/4$, while $\theta=\pi/2$ represents a fully classical version.}
	\label{fig:stdVariation}
\end{figure}

This exponential increase is characteristic of quantum gates arising from path cancellation, and is essentially the sign problem well-known in QMC \cite{Troyer2005,hangleiter2019easing}. To see this, consider a generalized Hadamard gate described by
\begin{equation}
	{U}_{\text{H,rotated}}=\kbordermatrix{
		&|0^{(d-1)}\rangle & |1^{(d-1)}\rangle  \\
		\langle 0^{(d)}| &\cos{\left(\theta\right)} & \sin{\left(\theta\right)} \\
		\langle 1^{(d)}| &\sin{\left(\theta\right)} & -\cos{\left(\theta\right)} }
	\label{eq:HadT}
\end{equation}
\noindent With $\theta=\pi/4$ we have the standard Hadamard gate, but with $\theta=\pi/2$ it becomes a classical gate with only positive weights.

We still have $2^d$ total paths, half of which go to $\vert 0\rangle$ and the other half to $\vert 1\rangle$. But the paths have different contributions (for $\theta\neq \pi/4$) and it takes more work to calculate the standard deviations, $\text{std}(\{W_C\},\{p_C\})$ analytically from (\ref{eq:U30}). Instead, we show numerical results in Fig.~\ref{fig:stdVariation}  for $n=1,\,d=10$ separately for paths that go to $\vert 0\rangle$ and for paths that go to $\vert 1\rangle$. In each case we show results for uniform sampling (equal $p_C$) and for non-uniform sampling based on \ref{eq:weightedS}.

The result for $\theta= \pi/4$ is the same as what we argued earlier in (\ref{eq:stdT}) for paths leading to $\vert 1\rangle$. Note how a combination of rotation ($\theta$) and non-uniform sampling ($p_{C}$) helps reduce the std and hence the noise down to zero in the classical limit. This is an elementary example of how the sign problem can be ``tamed" \cite{Marvian2019,hangleiter2019easing,Berg1606,hangleiter2019easing} and we mention it as an important option that can help make the probabilistic emulation of quantum circuits more efficient in the future.

\section{EXAMPLE 2: Multiple qubits, shallow circuit}
\label{sec:Shor}

\begin{figure*}[!ht]
	\centering
	\vspace{0pt}
	\includegraphics[width=0.75\linewidth,keepaspectratio]{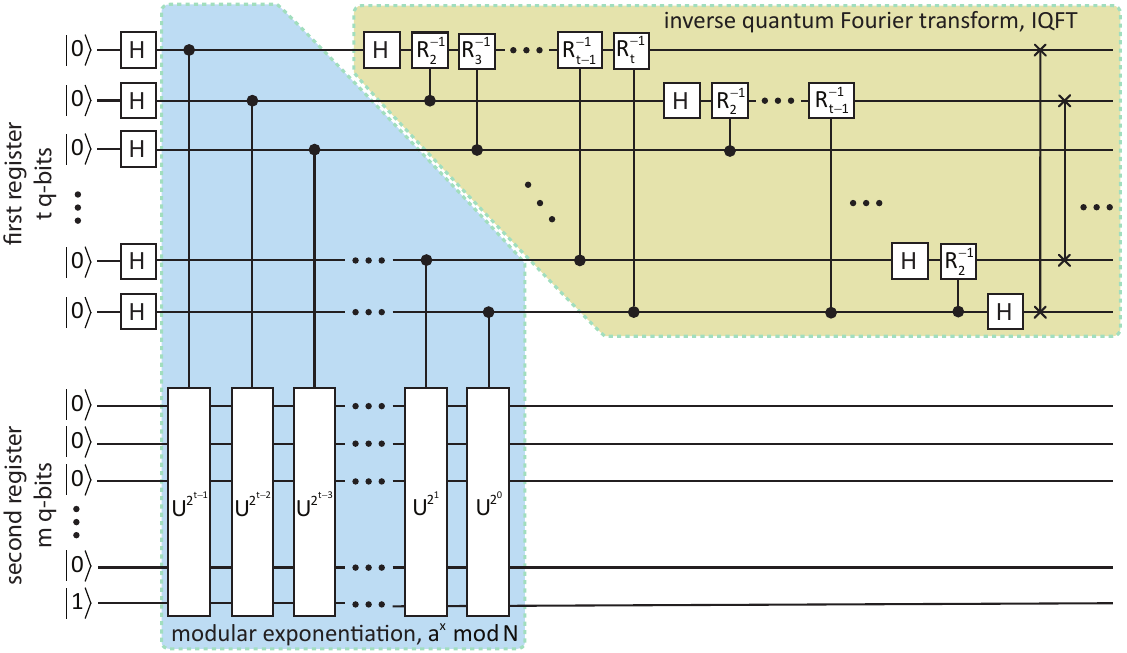}
	\caption{\textbf{Shor's order-finding circuit:} The circuit consists of two quantum registers: the first register is of size $t$ (satisfying $N^2\leq 2^t < 2N^2$) qubits and initially set to $\vert0\rangle$ state, while the second register is of size $n$ (satisfying $n= \left \lceil{ \log_2{N}}\right \rceil$) qubits and initially set to $\vert1\rangle$ state. First, a bank of Hadamard gates is applied to the top register to create a superposition of all classically possible states, $\sum_{x=0}^{2^t-1}\vert x\rangle$. Then a modular exponentiation operation, $f(x)=a^x \text{ mod } N$ is performed on the bottom register for each $\vert x\rangle$ in the superposed state. The ${U}$ transformation in the modular exponentiation block performs ${U}\vert u\rangle=\vert au \text{ mod } N\rangle$. After that, a $t$ qubit inverse quantum Fourier transform is performed which generates $r$ distinct peaks (which is the order of $f(x)$) in the probability distribution of the qubits in the top register.}
	\label{fig:Shorcircuit}
\end{figure*}

\begin{figure}[!ht]
	\vspace{0pt}
	\centering
	\includegraphics[width=0.9\linewidth,keepaspectratio]{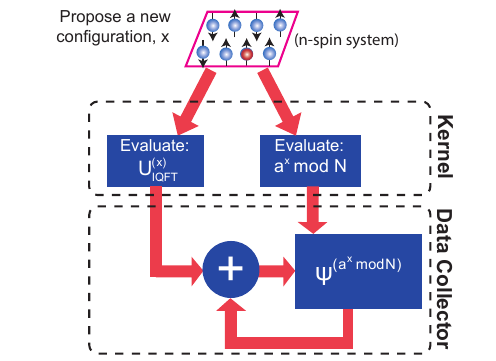}
	\caption{\textbf{Column sampling approach to period finding circuit}: shows the column sampling probabilistic approach used in this work. First, a set of $t=n$ spins corresponding to the qubits in the top register generates a random configuration $x$. This configuration is then used to evaluate the corresponding column of the IQFT ($U_{\text{IQFT}}^{(x)}$). A deterministic computation of $a^x \text{ mod } N$ is also performed. Instead of storing the output wavefunction ($\psi$) as a $2^{2n}\times 1$ column vector, it is stored in a $2^n \times 2^n$ matrix where each column corresponds to a value of $a^x \text{ mod } N$. The evaluated IQFT column, $U_{\text{IQFT}}^{(x)}$ is then added to the $\left(a^x \text{ mod } N\right)$\textsuperscript{th} column of $\psi$. This process is repeated until sufficient samples are taken. Finally, probabilities are calculated by taking the sum of absolute value squared along each row of $\psi$ which reduces to a $2^n\times1$ column vector.}
	\label{fig:ShorColumn}
\end{figure}

\begin{figure}[!ht]
	\centering
	\vspace{0pt}
	\includegraphics[width=0.95\linewidth]{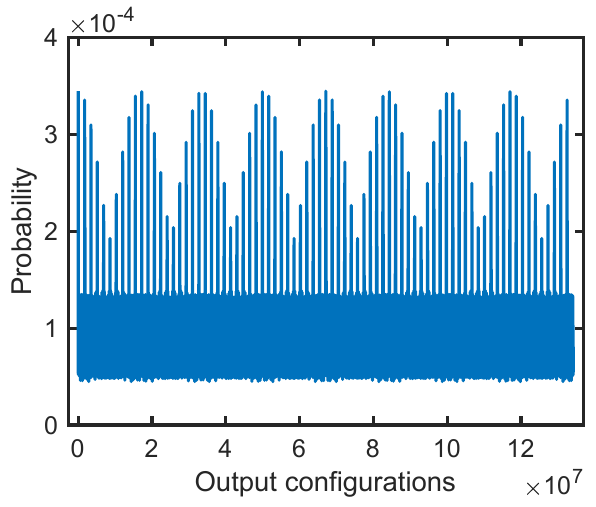}
	\caption{\textbf{Finding order of the function, $f(x)=10424^x \text{ mod } 98743069$}: the 78 distinct peaks indicates a period of $r = 78$. This figure is obtained with $250N$ samples, where the peaks can be clearly identified.}
	\label{fig:NewShorcircuit}
\end{figure}

Shor's algorithm for integer factorization is a cornerstone example that is often cited to refer to the computational advantage that can be harnessed from a quantum computer. Shor's algorithm ingeniously employs a quantum circuit to accelerate the process of finding the period $r$ of the function $f(x)=a^x \text{ mod }N$, where $N$ is the number to be factored and $a$ is a random number in the range $1 < a < N$ satisfying $\text{gcd}(a,N)=1$.

\subsection{Quantum circuit}
The commonly referred quantum circuit corresponding to this operation is shown in Fig.~\ref{fig:Shorcircuit}, employing two registers that are entangled to produce an output $\vert x, a^x \text{ mod } N \rangle$. A quantum Fourier Transform (QFT) is performed on the output of the first register to yield $\vert k, a^x \text{ mod } N \rangle$ on which measurements are made to obtain samples corresponding to one of the $r$ peaks, from which the value of $r$ is extracted using the continued fraction algorithm.  In order to ensure the high success rate of this algorithm with a few samples, the number of qubits $t$ in the first register is chosen such that $2n\leq t \leq 2n+1$, where $n$ is the minimum number of bits needed to represent the number $N$ being factorized.

\subsection{Probabilistic circuit}

We could map the quantum circuit in Fig.~\ref{fig:Shorcircuit} to a probabilistic circuit exactly. But it is possible to accelerate the implementation significantly with a few changes (Fig.~\ref{fig:ShorColumn}):

\begin{itemize} 
	\item We use only $t = n$ input p-bits in the first register, instead of the $2n\leq t \leq 2n+1$ qubits necessitated by the continued fraction algorithm. We obtain $r$ directly by counting the number of peaks in the output probability distribution.
	\item Each sample can consist of a family of paths, each leading to one output, rather than a single path; leading to a single output as shown in Fig.~\ref{fig:ShorColumn}. The idea is similar to the elementary version of \emph{column sampling} technique used in the field of randomized numerical linear algebra (randLNA) \cite{Drineas2016,Plancher2019}.
	\item The noise goes down $\sim$ the number of samples T, while the peak strength $\sim r$, so that we expect to need $T > r$ to identify the peaks correctly. Our numerical experiments suggest that in general one requires $T=\gamma r$ column samples, typically with $2\leq\gamma\leq 4$. The sampling process can be monitored regularly to decide when the peaks are clear and sampling can be stopped.  
	\item A straightforward implementation would require $2^n \times 2^n$ elements to be stored in order to keep track of $\vert k, a^x \text{ mod } N \rangle$. However, in our work, this memory requirement was reduced to $3 \times 2^n$ by first generating the random column numbers and evaluating $a^x \text{ mod } N$ operations for each $x$.
\end{itemize}

\noindent Using this probabilistic approach, we have factorized numbers up to $\sim 10^8$ as we will describe next.

\subsection{Performance Comparison}

Deterministic matrix multiplication schemes have been used to factorize numbers using the Shor algorithm, see for example \cite{Zulehner2019}. For comparison we choose an example implemented recently on a supercomputer \cite{Dang2019} and use our probabilistic approach to do the same problem on a laptop, and compare them in Table~\ref{tab:shor_comp}. We see that the probabilistic approach uses significantly less memory than the other approach mainly attributed to its reduction in the number of qubits in the first register.

\begin{table}[!ht]
	\centering
	\caption{Performance comparison between matrix product state approach implemented on a supercomputer (Ref.~\cite{Dang2019}) versus probabilistic approach used in this work for the factorization of a 20-bit number $N = 961307$, $a = 5$ and $r = 479568$.}
	\vspace{0.5cm}
	\begin{tabular}{@{}LCC@{}} \toprule
		\multirow{2}{*}{\parbox{2cm}{\textbf{Properties}}} & \multirow{2}{*}{\parbox{2cm}{Ref.~\cite{Dang2019}}} & \multirow{2}{*}{\parbox{1.8cm}{\textbf{Probabilistic approach}}}\\[1em] \midrule
		%\multirow{2}{*}{\parbox{2cm}{\# of bits in the integer, $n$}} & \multirow{2}{*}{\parbox{2cm}{20 }}& \multirow{2}{*}{\parbox{2cm}{20}}\\[2em]
        \multirow{2}{*}{\parbox{1.8cm}{\# of bits in the integer, $n$}} & \multirow{2}{*}{\parbox{0.3cm}{20 }}& \multirow{2}{*}{\parbox{0.3cm}{20}}\\[3em]
		\# of q-bits used & 60 & 40\\[0.5em]
		Time [s] & ~ 8 hours & 32 hours \\[0.5em]
		Memory  & 13.824 TB & $56$ MB \\[0.5em]
		Simulator & GNU compiler based & MATLAB based\\ [0.5em]
		\multirow{10}{*}{\parbox{2cm}{Machine
				description}} & \multirow{10}{*}{\parbox{3cm}{Magnus \cite{Magnus}, a Cray XC40 supercomputer
				with 24 cores at 2.60 GHz and 64 GB of RAM per
				node (implementation required total 216 nodes)}}
		& \multirow{10}{*}{\parbox{3cm}{64-bit  Laptop machine
				running at a clock frequency of
				2.6 GHz and 16 GB of memory running Windows 10.
				Used only a single core of the six available}}\\
		&&\\
		&&\\
		&&\\
		&&\\
		&&\\
		&&\\
		&&\\
		&&\\
		&&\\ \bottomrule
		
	\end{tabular}
	\label{tab:shor_comp}
\end{table}

Our improved probabilistic approach thus requires $O(r)$ column samples, with each column requiring the evaluation of $2^n$ elements which is of course, inferior to the polynomial scaling in $n$ expected from a qubit implementation that has made Shor's algorithm legendary and in general hard to emulate with classical resources \cite{Dang2019,tankasala2019,Lanyon2008,Monz1068,Vandersypen2001,nest2012efficient}. However, it does give significant improvement in memory requirement, and could also give significantly better computation time if implemented on an Ising machine with high sample throughput. 

Even on our single-core laptop, we are able to factorize much larger numbers than could be implemented deterministically. For example,
Fig.~\ref{fig:NewShorcircuit}, shows the output distribution obtained from the probabilistic approach when used to find the period of $f(x)=10424^x \text{ mod } 98743069$ (the period of $f(x)$ is $r=78$). We note that $98743069$ is a $27$-bit number and the period is obtained using only $250$ column samples out of $2^{27} \approx 134$ million columns on the aforementioned machine in less than 2 hours (in $\sim 5923$ s). But this advantage diminishes gradually as the period increases because the number of column samples required to detect peaks clearly also grows almost linearly with the period as mentioned earlier.

\section{Hardware emulation}
\label{sec:Hardware}

%Finally, we would like to draw attention to the possibility of \textit{hardware acceleration} over and above the \textit{algorithmic acceleration} discussed above.

The results in Table~\ref{tab:shor_comp} of the previous section for the probabilistic approach may look underwhelming. However, we would like to draw attention to the possibility of \textit{hardware acceleration} over and above the \textit{algorithmic acceleration} discussed above. The results presented here are all based on algorithms implemented on a general-purpose CPU/GPU architected like a general-purpose Turing machine. However, it is known that significant acceleration is possible using special-purpose processors in the form of a dedicated circuit. It was recently shown that a 3D Ising computer implementing an AQC algorithm for stoquastic problems could be accelerated by orders of magnitude using digital circuits implemented on FPGA, and even more with mixed signal circuits \cite{schowdhury2023}. Similar principles can be adapted to the GQC problems addressed in this paper. However, an additional unit for ``sign correction" is needed to keep track of the phase of each sample. This is in addition to the $sampling$ unit needed for stoquastic circuits for which the energy function has no imaginary part.

Both the Hadamard gate and the Quantum Fourier Transform circuit have purely imaginary energy functions. Since the real part is zero, samples are generated uniformly, all configurations having the same probability. Typically this would require us to collect $2^{nd}$ samples which is very challenging once $nd$ exceeds say $30$. However, if we can use suitable transformations like the rotated Hadamard gate to introduce a real part in the energy function then powerful sampling algorithms like Gibbs sampling can be harnessed to get acceptable results with far fewer samples and perhaps even escape the exponential scaling with $nd$.

\section{Conclusions}

In summary, we have presented a systematic and general procedure to translate any given quantum circuit exactly into a network of classical probabilistic p-bits each state of which is described by a complex amplitude. Although this mapping in general leads to the exponential asymptotic scaling with the size of the network and does not offer \emph{algorithmic} advantage, we expect a significant lowering of the prefactor by realizing that our general mapping makes it possible to directly transform any quantum circuit into a probabilistic network which then can be applied to a vast variety of fast-developing special-purpose Ising machines and thereby turning them into probabilistic simulators of any quantum circuit. This will also help in the growing effort to push the boundaries of the simulability of quantum circuits with classical/probabilistic resources and compare them with NISQ-era quantum computers.

\section{Appendix: Zero elements in U-matrix}
\label{App}

The procedure laid out in Sec.~\ref{sec:E} for obtaining energy functions for given transformation matrices is straightforward. One point to note, however, is that often there are zero elements in the matrix which will lead to singular values for the logarithmic functions in Eqs.~(\ref{eq:EB}) or~(\ref{eq:ET2}). A general but approximate way to deal with zero elements is to replace them with $e^{-J}$,  $J$ being a suitably large positive number. However, in special cases, an exact approach is possible, which is best illustrated with examples.

\textbf{One qubit examples:}  Consider the transformation  ${R}_z(\gamma)$ representing a one qubit rotation about the $z$-axis:
\begin{equation}
	{R}_z\left(\gamma\right)=e^{-i\gamma\sigma_z/2}=\kbordermatrix{
		&|0^{(d-1)}\rangle & |1^{(d-1)}\rangle\\
		\langle 0^{(d)}| & e^{-i \gamma/2} & 0 \\
		\langle 1^{(d)}| & 0 & e^{+i \gamma/2}}
\end{equation}
Replacing the zero off-diagonal elements with $e^{-J}$, $J$ being a large positive number, we have
\begin{eqnarray}
	E^{(d)} &=& -i \frac{\gamma}{2} \big( 1- s^{(d-1)} - s^{(d)} \big) \nonumber\\
	&+& \lim\limits_{J\to\infty} {J \big(s^{(d-1)} + s^{(d)} -  2s^{(d-1)}s^{(d)} \big)}
\end{eqnarray}
This is the straightforward approximate approach described above. Alternatively, we could eliminate $s^{(d)}$ as an independent variable by setting it equal to $s^{(d-1)}$, so that
\begin{eqnarray}
	&s^{(d)} = s^{(d-1)} \nonumber \\
	&E^{(d)} = -i \frac{\gamma}{2} \big( 1 - 2 s^{(d-1)}  \big)
\end{eqnarray}

This approach can also be used for zeros on the diagonal. Consider the Pauli-X gate:
\begin{equation}
	{X} =  \ \left[
	\begin{array}{r r r}
		0 & 1\\
		1 & \ 0 \\
	\end{array}
	\right] 
\end{equation}
\noindent We can eliminate $s^{(d)}$ as an independent variable by setting it equal to $1-s^{(d-1)}$, so that
\begin{eqnarray}
	& s^{(d)} = 1-s^{(d-1)} \nonumber \\
	&E^{(d)} = 0
\end{eqnarray}
\noindent Note that this is essentially a deterministic NOT gate.

\textbf{Two qubit exchange operator:} Consider the two qubit exchange operator acting between qubit $1$ and $2$ defined as follows:
\begin{equation}
	{J}_{12}=\kbordermatrix{
		&|00^{(d-1)}\rangle & |10^{(d-1)}\rangle & |01^{(d-1)}\rangle & |11^{(d-1)}\rangle\\
		\langle 00^{(d)}| &e^{i \gamma /2} &0 &0 &0\\
		\langle 10^{(d)}| &0 &e^{-i \gamma / 2} &0 &0 \\
		\langle 01^{(d)}| &0 &0 &e^{-i \gamma / 2} &0 \\
		\langle 11^{(d)}| &0 &0 &0 &e^{i \gamma /2} }
\end{equation}
\vspace{0.05in}

\noindent Since the matrix is diagonal, we can eliminate $s_1^{(d)}, s_2^{(d)}$ by writing

\begin{eqnarray}
	&s_1^{(d)} = s_1^{(d-1)} \ , \  s_2^{(d)} = s_2^{(d-1)} \nonumber \\
	&E^{(d)} = \frac{i \gamma}{2} \big( 1 - 2 s_1^{(d-1) }  \big) \big( 1 - 2 s_2^{(d-1)}  \big)
\end{eqnarray}

\noindent For two-qubit operations with off-diagonal elements, the general approach would be to use $\exp{(-J)}$ to replace zero elements if present. Instead, we may be able to reduce the number of independent variables for specific cases as illustrated below.

\textbf{CNOT Gate:} Consider the CNOT gate described by the transformation matrix:
\vspace{-0.05 in}
\begin{equation}
	{U}_{\text{CNOT}}=\kbordermatrix{
		&|00^{(d-1)}\rangle & |10^{(d-1)}\rangle & |01^{(d-1)}\rangle & |11^{(d-1)}\rangle\\
		\langle 00^{(d)}| &1 &0 &0 &0\\
		\langle 10^{(d)}| &0 &1&0 &0 \\
		\langle 01^{(d)}| &0 &0 &0 &1 \\
		\langle 11^{(d)}| &0 &0 &1 &0 }\nonumber
\end{equation}
\noindent In this case, we can write
\begin{eqnarray}
	& s_1^{(d)} = s_1^{(d-1)}+s_2^{(d-1)} -2 s_1^{(d-1)}s_2^{(d-1)}\nonumber \\
	& s_2^{(d)} = s_2^{(d-1)} \nonumber \\
	&E^{(d)} = 0
\end{eqnarray}
\noindent It can be seen that this is essentially a deterministic XOR gate:
\begin{equation}
	s_1^{(d)} =  \mathrm{XOR} \  \big(s_1^{(d-1)} , s_2^{(d-1)} \big)
\end{equation}

\textbf{CCNOT Gate:} Similarly for the CCNOT gate, we can write
\begin{eqnarray}
	\footnotesize
	& s_1^{(d)} = s_1^{(d-1)}+s_2^{(d-1)} s_3^{(d-1)} -2 s_1^{(d-1)}s_2^{(d-1)} s_3^{(d-1)} \nonumber \\
	& s_2^{(d)} = s_2^{(d-1)}, s_3^{(d)} = s_3^{(d-1)} \nonumber \\
	&E^{(d)} = 0
\end{eqnarray}
\normalsize \noindent This too is essentially a deterministic gate:
\begin{equation}
	s_1^{(d)} =  \mathrm{XOR} \  \bigg( s_1^{(d-1)} , \  \big( s_2^{(d-1)}  \ \mathrm{AND} \ s_3^{(d-1)} \big) \bigg)
\end{equation}

\vspace{5pt}
\section*{Acknowledgment}
This work was supported in part by ASCENT, one of six centers in JUMP, a Semiconductor Research Corporation (SRC) program sponsored by DARPA, in part by Center for Science of Information (CSoI), an NSF Science and Technology Center, under grant CCF-0939370 and in part by Office of the Naval Research YIP program. The authors thank Brian Sutton and Marc Cahay for extensive comments and suggestions on an earlier version of this manuscript. The authors are also grateful to Diptiman Sen for useful discussions related to non-stoquastic Hamiltonians. The authors also thank Jan Kaiser for many helpful discussions, especially those related to variance estimations. The authors also thank Dr. Seokmin Hong for helpful discussions regarding Fig. 8. S. C. was with Elmore Family School of Electrical and Computer Engineering, Purdue University, IN 47907, USA when this work was done.

\vspace{5pt}
\section*{Conflict of Interest}
Supriyo Datta has a financial interest in Ludwig Computing. The authors declare no other competing interests.

\vspace{10pt}
\footnotesize

\end{document}